\documentstyle[preprint,prc,aps,floats,epsf]{revtex}

\tightenlines

\begin{document}
%
%
\def\symdef#1#2{\def#1{#2}}
\symdef{\alphabar}{\overline\alpha}
\symdef{\alphabarp}{\overline\alpha\,{}'}
\symdef{\Azero}{A_0}

\symdef{\bzero}{b_0}

\symdef{\Cs}{C_{\rm s}}
\symdef{\cs}{c_{\rm s}}
\symdef{\Cv}{C_{\rm v}}
\symdef{\cv}{c_{\rm v}}

\symdef\dalem{\frame{\phantom{\rule{8pt}{8pt}}}}
\symdef{\Deltaevac}{\Delta{\cal E}_{\rm vac}}

\symdef{\ed}{{\cal E}}
\symdef{\edk}{{\cal E}_k}
\symdef{\edkzero}{{\cal E}_{k0}}
\symdef{\edv}{{\cal E}_{\rm v}}
\symdef{\edvphi}{{\cal E}_{{\rm v}\Phi}}
\symdef{\edvphizero}{{\cal E}_{{\rm v}\Phi 0}}
\symdef{\edzero}{{\cal E}_{0}}
\symdef{\Efermistar}{E_{{\scriptscriptstyle \rm F}}^\ast}
\symdef{\Efermistarzero}{E_{{\scriptscriptstyle \rm F}0}^\ast}
\symdef{\etabar}{\overline\eta}
\symdef{\ezero}{e_0}

\symdef{\fomega}{f_\omegav}
\symdef{\fpi}{f_\pi}
\symdef{\fv}{f_{\rm v}}
\symdef{\fvt}{\widetilde\fv}

\symdef{\gA}{g_A}
\symdef{\gammazero}{\gamma_0}
\symdef{\gomega}{g_\omegav}
\symdef{\gpi}{g_\pi}
\symdef{\grho}{g_\rho}
\symdef{\gs}{g_{\rm s}}
\symdef{\gv}{g_{\rm v}}

\symdef{\fm}{\mbox{\,fm}}

\symdef{\hDirac}{h_0}
\symdef{\hTDirac}{h_1}
\symdef{\hLSFW}{h_{{\scriptscriptstyle\rm LS,FW}}}
\symdef{\hLSR}{h_{{\scriptscriptstyle\rm LS,R}}}
\symdef{\hLST}{h_{{\scriptscriptstyle\rm LS,T}}}
\symdef{\hNR}{h_{{\scriptscriptstyle\rm NR}}}

\symdef{\infm}{\mbox{\,fm$^{-1}$}}

\symdef{\kappabar}{\overline\kappa}
\symdef{\kfermi}{k_{{\scriptscriptstyle \rm F}}}
\symdef{\kfermizero}{k_{{\scriptscriptstyle \rm F}0}}
\symdef{\Kzero}{K_0}

\symdef{\lambdabar}{\overline\lambda}
\symdef{\LdotS}{\bbox{\sigma\cdot L}}
\symdef{\lzero}{l_{0}}

\symdef{\Mbar}{\overline M}
\symdef{\Mbarzero}{\Mbar_0}
\symdef{\MeV}{\mbox{\,MeV}}
\symdef{\momega}{m_\omegav}
\symdef{\mpi}{m_\pi}
\symdef{\mrho}{m_\rho}
\symdef{\ms}{m_{\rm s}}
\symdef{\Mstar}{M^\ast}
\symdef{\Mstarzero}{M^\ast_0}
\symdef{\mv}{m_{\rm v}}
\symdef{\mzero}{{\rm v}_{0}}

\symdef{\Nbar}{\overline N}

\symdef{\omegaV}{V}
\symdef{\omegav}{{\rm v}}

\symdef{\Phizero}{\Phi_0}
\symdef{\psibar}{\overline\psi}
\symdef{\psidagger}{\psi^\dagger}
\symdef{\pvec}{{\bf p}}

\symdef{\rhoB}{\rho_{{\scriptscriptstyle \rm B}}}
\symdef{\rhoBzero}{\rho_{{\scriptscriptstyle \rm B}0}}
\symdef{\rhos}{\rho_{{\scriptstyle \rm s}}}
\symdef{\rhospzero}{\rho'_{{\scriptstyle {\rm s} 0}}}
\symdef{\rhoszero}{\rho_{{\scriptstyle {\rm s}0}}}
\symdef{\rhozero}{\rho_0}

\symdef{\Szero}{S_0}

\symdef{\Tr}{{\rm Tr\,}}

\symdef{\umu}{u^\mu}
\symdef{\Ualpha}{U_{\alpha}}
\symdef{\Ueff}{U_{\rm eff}}
\symdef{\Uzero}{U_0}
\symdef{\Uzerop}{U_0'}
\symdef{\Uzeropp}{U_0''}

\symdef{\vecalpha}{{\bbox{\alpha}}}
\symdef{\veccdot}{{\bbox{\cdot}}}
\symdef{\vecnabla}{{\bbox{\nabla}}}
\symdef{\vecpi}{{\bbox{\pi}}}
\symdef{\vectau}{{\bbox{\tau}}}
\symdef{\vecx}{{\bf x}}
\symdef{\Vopt}{V_{\rm opt}}
\symdef{\Vzero}{V_0}

\symdef{\wt}{\widetilde}
\symdef{\wzero}{w_0}
\symdef{\Wzero}{W_0}

\symdef{\zetabar}{\overline\zeta}
%
%
%
%

\draft

\preprint{IU/NTC\ \ 97--09}

\title{The Nuclear Spin-Orbit Force in\\
       Chiral Effective Field Theories}

\author{R. J. Furnstahl and John J. Rusnak}
\address{Department of Physics \\
         The Ohio State University,\ \ Columbus, OH\ \ 43210}
\author{Brian D. Serot}
\address{Department of Physics and Nuclear Theory Center \\
         Indiana University,\ \ Bloomington, IN\ \ 47405}
%
%

%
\date{September, 1997}

\maketitle
\begin{abstract}
A compelling feature of relativistic mean-field phenomenology
has been the reproduction of spin-orbit splittings in finite nuclei 
after fitting only to equilibrium properties of infinite nuclear matter.
This successful result occurs when the velocity dependence
of the equivalent central potential that leads to saturation arises primarily
because of a reduced nucleon effective
mass.
The
spin-orbit interaction is then also specified
when one works in a four-component Dirac framework.
Here the nature of the spin-orbit force in more general chiral
effective field theories of nuclei is examined,
with an
emphasis on the role of the tensor coupling of the isoscalar
vector meson ($\omega$) to the nucleon.

\end{abstract}

\section{Introduction}

The concepts and methods of effective field theory (EFT)
\cite{WEINBERG79,LEPAGE89,POLCHINSKI92,GEORGI93,WEINBERG95}
have recently elucidated the 
successful nuclear phenomenology of relativistic field theories
of hadrons, called quantum hadrodynamics (QHD) 
\cite{SEROT86,SEROT92,FURNSTAHL97,SEROT97}.
The EFT framework shows how QHD models can be consistent with
the symmetries of quantum chromodynamics (QCD) 
and can be extended to accurately reproduce its low-energy features.
The EFT perspective 
accounts for the success of relativistic mean-field 
models and provides an
expansion scheme at the mean-field level and
for going beyond it \cite{FURNSTAHL97,RUSNAK97}.

One of the most compelling features of QHD mean-field phenomenology
is the reproduction of spin-orbit splittings in finite nuclei 
after fitting only to the equilibrium properties of infinite nuclear matter.
This successful result occurs when the velocity dependence
of the equivalent central potential that leads to saturation arises primarily
because of a reduced nucleon effective mass.
The spin-orbit interaction in nuclei is then also specified
by the structure of the four-component Dirac spinors \cite{SEROT97}.
Does this connection survive in the extended EFT framework?
Here we re-examine the nature of the spin-orbit force in chiral
effective field theories of nuclei.

The strong correlation in QHD models
between the effective mass $\Mstarzero$ of the nucleon 
in equilibrium nuclear matter and the strength of the
spin-orbit force in nuclei has been well documented
\cite{REINHARD89,GAMBHIR90,BODMER91,SEROT97}.
Typical values of $\Mstarzero/M$ in successful models are roughly 0.6, which
yield spin-orbit splittings close to experimental values
(e.g., 6\,MeV for $p$ states in $^{16}$O).
In Ref.~\cite{FURNSTAHL95}, a wider class of models,
which included variations of the linear sigma model as well as
all of the standard QHD models used phenomenologically,%
\footnote{These ``conventional'' QHD models typically 
include cubic and quartic self-couplings of the scalar field and sometimes
a quartic self-coupling of the vector field, and they may also include
one-loop vacuum corrections.} 
was studied systematically.
Accurate reproductions of observables in finite nuclei, including the
spin-orbit splitting in particular, 
tightly constrained the value of $\Mstarzero/M$
to the range $0.58 \leq \Mstarzero/M \leq 0.64$.

In Ref.~\cite{FURNSTAHL97}, an effective hadronic lagrangian
consistent with the symmetries of QCD and intended for application
to finite density systems was constructed.
This involved a different strategy than in conventional QHD
phenomenology.
Rather than maximizing predictivity by minimizing the free parameters
(as in renormalizable models),
the goal was to test a systematic expansion
for low-energy observables,
which included the effects of hadron compositeness and the constraints
of chiral symmetry.
The degrees of freedom are (valence) nucleons, pions, and the
low-lying non-Goldstone bosons.
A scalar-isoscalar field with a mass of roughly 500\,MeV was also included
to simulate the exchange between nucleons of two correlated pions in this
channel.
The lagrangian was expanded in powers of the fields and their
derivatives, with the terms organized using Georgi's ``naive
dimensional analysis'' (NDA) \cite{GEORGI84b,GEORGI93,FRIAR96}.
The coefficient of each term is written as a combination of the
pion-decay constant $\fpi \approx 93\,$MeV and a larger scale
$0.5 \alt \Lambda \alt 1\,$GeV times a dimensionless coupling
constant.
The effective lagrangian is said to be ``natural'' if these dimensionless
coefficients are of order unity.

The result is
a faithful representation of low-energy, non-strange QCD, as long as all 
nonredundant terms consistent with symmetries are included.
In addition, the mean-field framework provides a means of
approximately including  higher-order many-body and loop effects,
since the scalar and vector meson fields play the role of
auxiliary Kohn--Sham potentials in relativistic density functional 
theory \cite{SEROT97}.
Fits to nuclear properties at the mean-field level
showed that the effective lagrangian could be truncated at the first
few powers of the fields and their derivatives,
with natural [$O(1)$] coefficients for each term.
Of the two full parameter sets identified, one 
(model G2) provided an excellent fit, including spin-orbit splittings,
with a nuclear matter value of $\Mstarzero/M \approx 0.66$.
More recently, new parameters sets with excellent fits have been found
with $\Mstarzero/M \approx 0.70$ \cite{RUSNAK97b}.

An analogous study was made of relativistic ``point-coupling'' (PC) models in 
Ref.~\cite{RUSNAK97}.
In these models, non-Goldstone mesons are not included explicitly.
One instead has an expansion of the nucleon scalar and vector potentials
in powers and derivatives of local nucleon scalar and vector densities.
As in Ref.~\cite{FURNSTAHL97}, an effective lagrangian
consistent with chiral symmetry was constructed and fits to nuclear
observables were made to test naturalness.
Excellent fits were found with unprecedented values of $\Mstarzero/M$,
as high as $0.74$ at nuclear matter equilibrium.

Our purpose here is to examine the nature of the nuclear
spin-orbit force in these generalized models.
We find that 
the most important new feature affecting spin-orbit splittings
in the lagrangians
of Refs.~\cite{FURNSTAHL97} and \cite{RUSNAK97} 
is the inclusion of a {\em tensor\/} coupling
of the omega field to the nucleon (or its point-coupling analog).
Because the effective lagrangian includes a derivative expansion,
the tensor term appears at a relatively low order.
None of the other new terms has a comparable impact on the spin-orbit
strength.

We find a trade-off between the size of the scalar potential
(or equivalently, the effective mass $\Mstar$) and
the size of the tensor coupling.
The existence of such a trade-off has been
noted in  previous studies of the isoscalar tensor coupling in QHD models,
but good fits to nuclei
were found only with relatively small values
of the coupling and thus small 
$\Mstarzero \approx 0.6 M$ \cite{RUFA88,REINHARD89}.
Furthermore, this coupling is usually taken to be zero in one-boson-exchange
potentials \cite{MACHLEIDT}
or limited to small values due to constraints from {\it free\/}
nucleon form factors and the assumption of vector 
meson dominance \cite{HOHLER76,BROWN86,MEISSNER87,MEISSNER}.
However, as an effective coupling in nuclei,
which might absorb higher-order many-body effects at the mean-field level, 
the isoscalar tensor coupling could be much larger than these constraints
dictate and still be of natural size.

In fits made using more complete meson and point-coupling models,
larger (but still natural) values are favored, 
particularly in the point-coupling models \cite{RUSNAK97}.
In this work,
we show that the increase in $\Mstarzero/M$ in these models is well
accounted for by the contribution from the tensor coupling.
First, we consider two-component reductions
of the models, which are reviewed in Sects.~\ref{sect:two} and \ref{sect:three}. 
The spin-orbit potential in the two-component single-particle
hamiltonian manifests the influence of both $\Mstarzero/M$ and the
tensor coupling, and
a simple local-density approximation leads to quantitative predictions
for the spin-orbit strength in nuclei.
We find a strong correlation between the size of the isoscalar tensor
coupling and the increase of $\Mstarzero / M$ above 0.6.
This discussion applies to a very general class of QHD models, including the 
quark-meson-coupling (QMC) model \cite{SAITO96} and the
Zim\'anyi-Moszkowski (ZM)
model \cite{BIRO97} (both of which are found to be deficient).

A second approach is to transform the nucleon fields in models
with tensor couplings.
A feature of effective lagrangians is the freedom to perform field
redefinitions without changing the physics
(if all nonredundant terms have been included).
It is not possible to completely transform away the tensor interaction in a
QHD effective lagrangian.
To study the consequences for nucleon--nucleus scattering,
however,  we can perform an energy-dependent transformation to
replace the potential in the Dirac equation with an equivalent
potential (that is, one that generates the same scattering observables) with
no tensor term \cite{CLARK85}.
The new scalar and vector potentials can then be compared to
empirical optical potentials.

It is well known that in
relativistic models applied to hypernuclei, the isoscalar tensor
coupling of the omega to the hyperon plays a critical role in
{\it reducing\/} the strength of the hyperon spin-orbit 
potential \cite{JENNINGS90,COHEN91}.
Quark-model arguments imply that the anomalous coupling is equal to
(minus) the vector coupling, which leads to a
strong cancellation between contributions to the spin-orbit force and
therefore small splittings, as needed phenomenologically.
To our knowledge, there is no corresponding implication for the
tensor coupling of the omega to nucleons.\footnote{%
We note, however, that in the SU(2) limit of a simple constituent quark
model, in which the quarks have only Dirac magnetic moments and couple
directly to the photon, the nucleon isoscalar magnetic moment is
small \protect\cite{DGH92}.}

\section{Spin-orbit Force and $\Mstar$ Without Tensor Coupling}
\label{sect:two}

In this section, we explore the relationship between the spin-orbit
force and the nucleon potentials in models without tensor potentials.
The connection between the spin-orbit splittings and the scalar and vector
potentials is most easily seen by making a reduction to
a two-component spin-orbit hamiltonian.
We can write the isoscalar part of the Dirac single-particle
hamiltonian (for spherical nuclei) as
\begin{equation}
   \hDirac = -i\bbox{\nabla\cdot\alpha} + 
        \beta \Bigl(M-\Phi(r)\Bigr)
      +  W(r)  \ , \label{eq:hdirac}
\end{equation}
where $M$ is the nucleon mass,
$\Phi$ is the scalar potential, and $W$ is the vector potential.
A tensor potential is considered in the next section.
Note that we need not assume that $\Phi$ is simply proportional
to a scalar meson field $\phi$.  
In fact, $\Phi$ could be proportional to $\phi$
(as in conventional QHD models), or could be expressed as a sum of scalar and
vector densities (as in relativistic point-couplings models), or could be
a nonlinear function of $\phi$ (as in the QMC and ZM models).

To leading-order in $1/M$, the
spin-orbit term in a Foldy--Wouthuysen reduction
of $\hDirac$
takes the form \cite{SEROT86}
\begin{equation}
   \hLSFW = \biggl[
   {1\over 4M^{2} r} 
        \left( {d \Phi \over dr} +  {d W \over dr} \right)
   + O(1/M^3) \biggr] \LdotS  \   \ .
     \label{eq:hFW}
\end{equation}
The expansion is in powers of momenta (i.e., derivatives) 
and potentials divided by the nucleon mass.

The qualitative impact of the potentials on the spin-orbit force is clear
from this expression and leads to some basic observations common to all
successful models:
\begin{enumerate}
  \item Successful models are natural \cite{FURNSTAHL97,RUSNAK97},
    which implies that the
   potentials closely follow the baryon density.
     This result follows from the equations for the potentials and the close
     relationship between scalar and vector densities,
\begin{eqnarray}
     \Phi(r) &=& {\alpha_s\over\fpi^{2}}\rhos(r)
            +   O \left({\Phi^{2}\over M^2},{W^{2}\over M^2},
                  {\nabla^{2}\Phi \over M^3}  \right)  
               \ , \label{eq:Phi} \\
     W(r) &=& {\alpha_v\over\fpi^{2}}\rhoB(r)
            +   O \left({\Phi W\over M^2},{\nabla^{2}W \over M^3} \right)  
               \ ,  \label{eq:W}
\end{eqnarray}
where $\alpha_s$ and $\alpha_v$ are $O(1)$.
The precise expressions for the constants $\alpha_s$ and $\alpha_v$
will depend on the model; for example, in conventional QHD models
$\alpha_s = \gs^2 \fpi^2/\ms^2$ and $\alpha_v = \gv^2 \fpi^2/\mv^2$.
In every case, however, 
naturalness implies that the corrections are order $\Phi/M$ 
$\approx 1/3$ smaller than the leading term in the interior 
of nuclei and smaller still on the surface.
Furthermore, the scalar density and baryon density are equal to within
5 to 10\% at ordinary nuclear densities.
  \item  
Thus a simple local density approximation for $\Phi(r)$
     and $W(r)$ should be quite reasonable:
   \begin{eqnarray}
     \Phi(r) &\approx& \Phizero {\rhoB(r)\over \rhozero} 
                   \ , \label{eq:Philda}  \\
     W(r) &\approx& \Wzero {\rhoB(r)\over \rhozero} 
                   \ , \label{eq:Wlda}   
   \end{eqnarray}
    where $\Phizero$ and $\Wzero$ are the
     values of the potentials in nuclear matter at equilibrium density
     $\rhozero$.
     The remaining factor from the expectation value of $\hLSFW$
     will be universal for different models, 
     as long as they accurately reproduce nuclear density profiles.
     \item 
     The Hugenholtz--van Hove theorem relates $\Phizero$ and $\Wzero$
     through the values of the chemical potential $\mu$ and the
     equilibrium Fermi momentum $\kfermi$:
 \begin{equation}
    \mu = \Wzero + \sqrt{\kfermi^{2} + \Mstarzero{}^{2}}  \ ,
       \label{eq:HvH} 
 \end{equation}
    with $\Mstarzero \equiv M - \Phizero$.
    Since $\mu$ and $\kfermi$ are the same in all successful models
    to very good approximation \cite{FURNSTAHL87,FURNSTAHL97}, 
    the dependence on $\Wzero$ can be
    eliminated in favor of $\Phizero$.
     \item Therefore, the strength of the spin-orbit
     interaction (\ref{eq:hFW}) and 
     the resulting magnitude of the splitting in nuclei should be,
     to a good approximation, directly 
     proportional to the
     size of $\Phizero$,
     the scalar potential in nuclear matter.
  \end{enumerate}

\begin{figure}[p]
 \setlength{\epsfxsize}{4.1in}
  \centerline{\epsffile{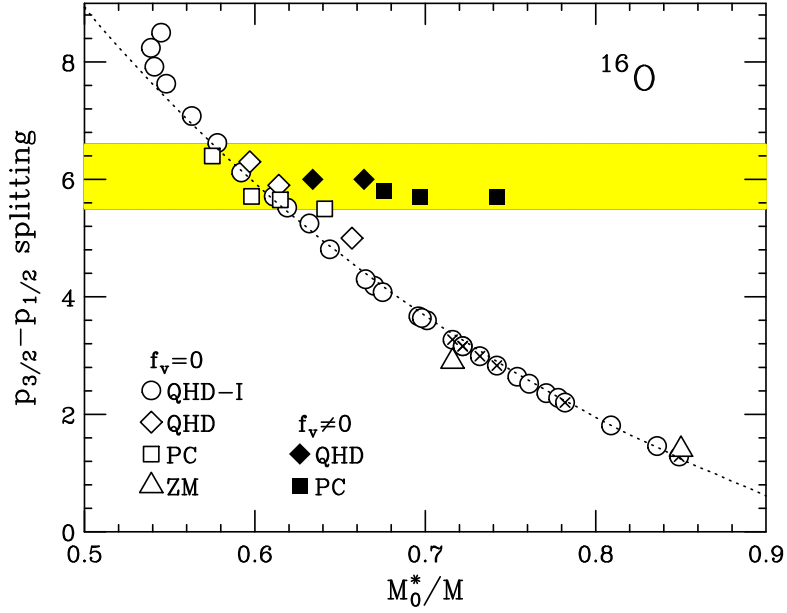}}
\vspace{.1in} 
\caption{\small Spin-orbit splitting for the proton $p$ states in
${}^{16}$O vs.\ equilibrium $\Mstarzero/M$ 
for a variety of models. 
The shaded band is an estimate of the experimental uncertainty.
Open symbols are for models with tensor coupling
$\fv=0$ and filled symbols are for
models with $\fv\neq 0$.
The dotted curve follows from Eq.~(\protect\ref{eq:split}).
See the text for descriptions of the models. }
 \label{fig:oxy}
\end{figure}

\begin{figure}[p]
 \setlength{\epsfxsize}{4.1in}
  \centerline{\epsffile{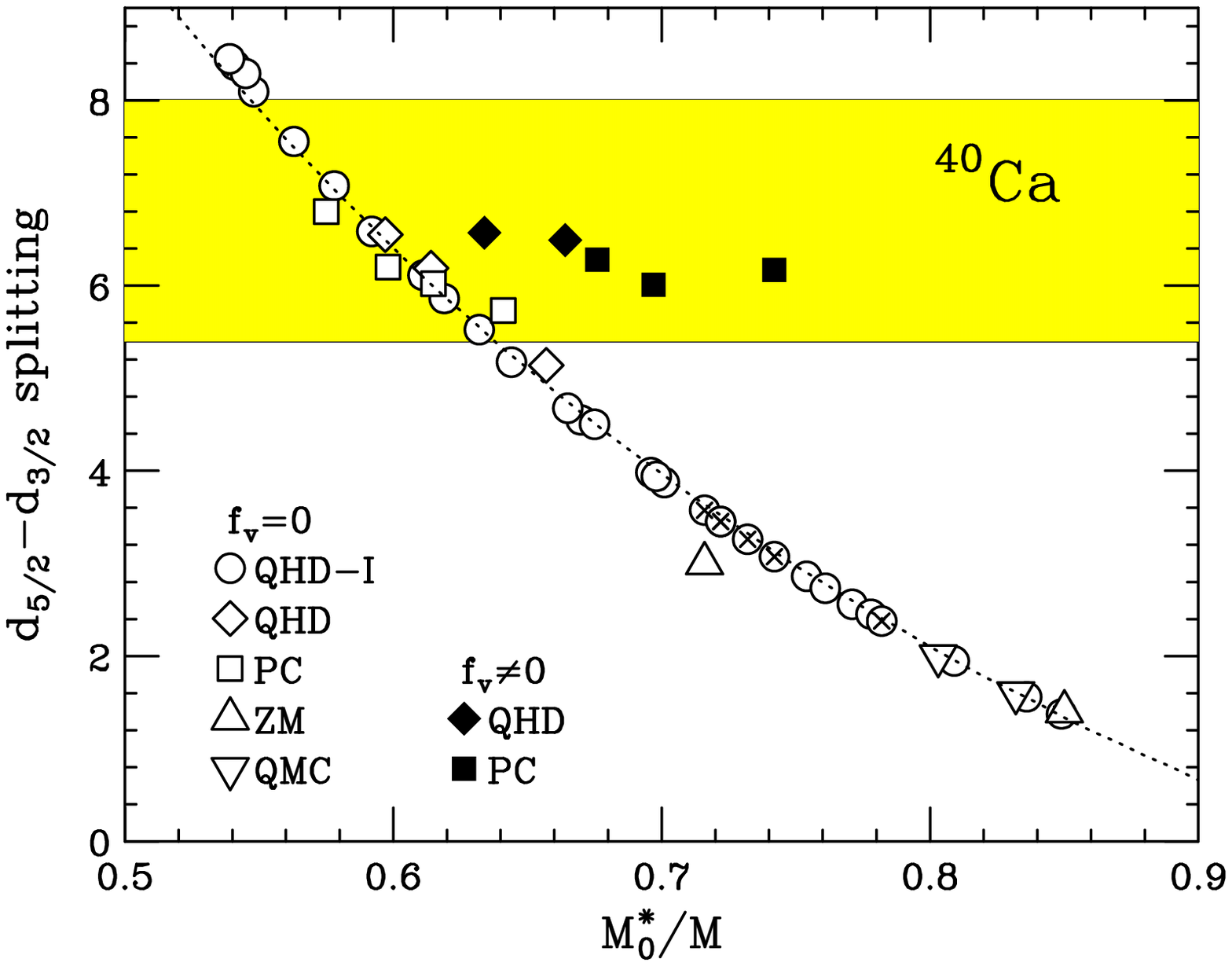}}
\vspace{.1in} 
\caption{\small Same as Fig.~\protect\ref{fig:oxy} for
the proton $d$ states in
${}^{40}$Ca.}
 \label{fig:cad}
\end{figure}

\begin{figure}[p]
 \setlength{\epsfxsize}{4.1in}
  \centerline{\epsffile{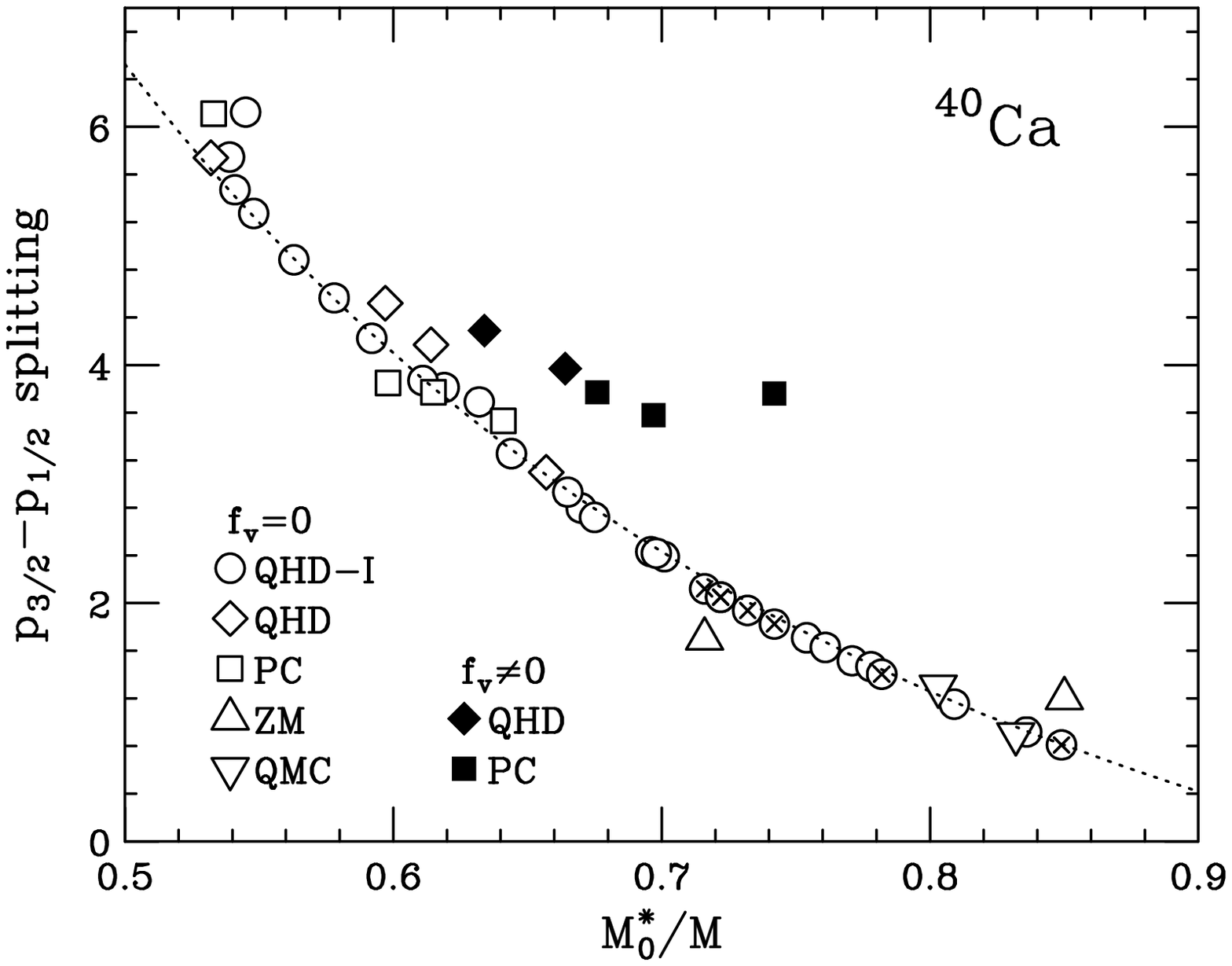}}
\vspace{.1in} 
\caption{\small Same as Fig.~\protect\ref{fig:oxy}
for the proton $p$ states in
${}^{40}$Ca.}
 \label{fig:cap}
\end{figure}

\begin{figure}[p]
 \setlength{\epsfxsize}{4.1in}
  \centerline{\epsffile{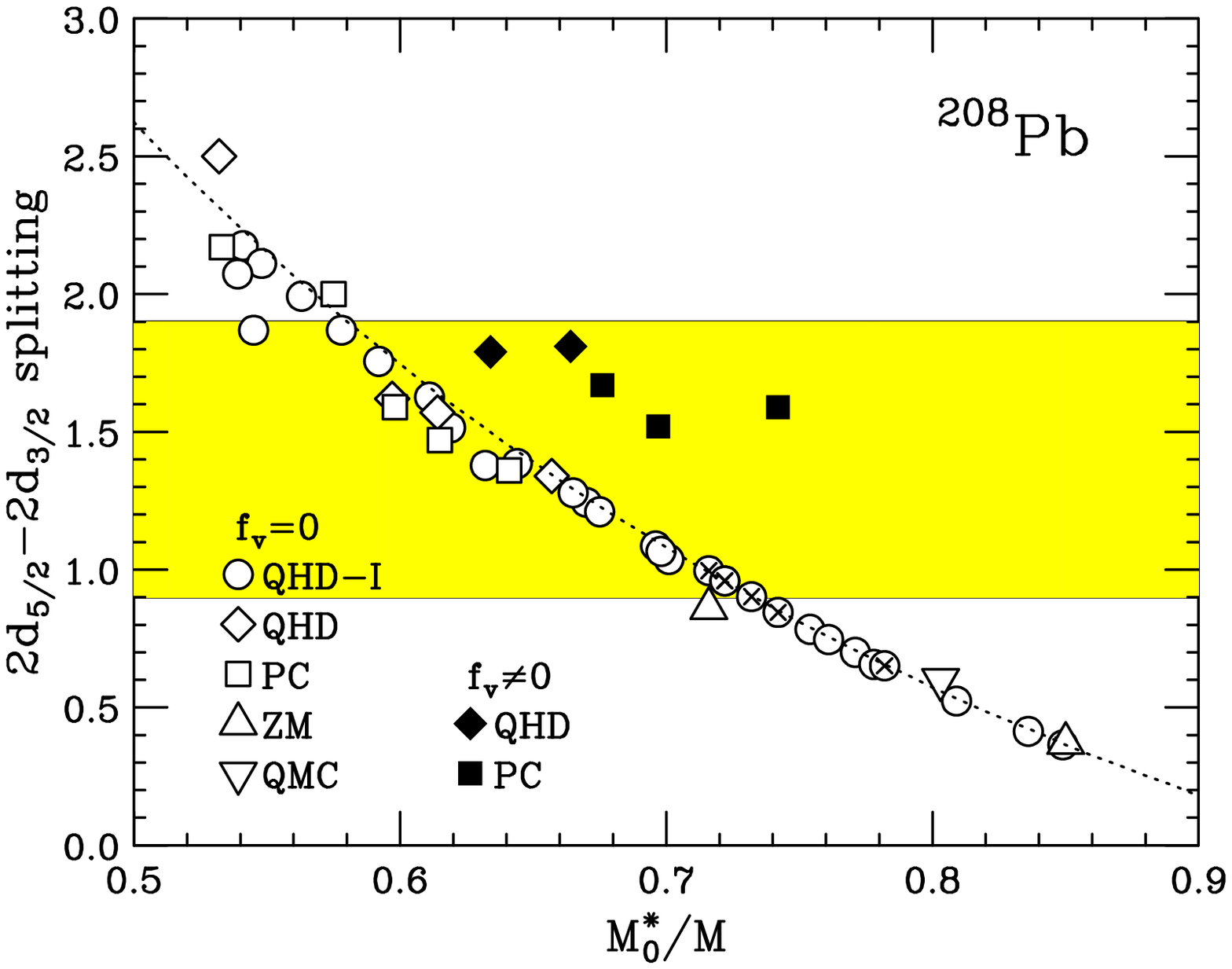}}
\vspace{.1in} 
\caption{\small Same as Fig.~\protect\ref{fig:oxy} 
for the proton $2d$ states in
${}^{208}$Pb.}
 \label{fig:pb}
\end{figure}

If this analysis is quantitatively valid, we would expect to find
that plots of spin-orbit splittings in nuclei versus equilibrium
$\Mstarzero/M$ for different models that reproduce nuclear density
distributions would lie roughly on a straight line.
Such plots are shown in Figs.~\ref{fig:oxy}--\ref{fig:pb} for splittings
in $^{16}$O, $^{40}$Ca, and $^{208}$Pb using a large
number of models. 
The open circles are  for QHD--I models \cite{FURNSTAHL87,SEROT97}
(the original linear Walecka model
plus cubic and quartic scalar self-interactions).
The circles marked with crosses also include one-loop vacuum corrections;
this is essentially equivalent to adding a particular quintic
scalar self-interaction term \cite{FURNSTAHL97b}.
Each QHD--I model was fit to a standard set of nuclear matter saturation
conditions (equilibrium density, binding energy, and symmetry energy)
and the charge radius of $^{40}$Ca \cite{FURNSTAHL87}.
The diamonds are for QHD models from Ref.~\cite{FURNSTAHL97},
squares are for point-coupling models from Refs.~\cite{NIKOLAUS92} 
and \cite{RUSNAK97},
and triangles are for QMC and ZM models from Refs.~\cite{SAITO96} and
\cite{CHIAP97}.
(The models with filled symbols
contain tensor interactions and are discussed in Sect.~\ref{sect:three}.) 
It is evident that there is a universal curve, but it is not
a straight line except for $\Mstarzero$ very close to $M$ (i.e., small
$\Phi$), in contrast to the conclusion reached in point (4), above.

For a more {\it quantitative\/} local-density estimate of the spin-orbit
splitting, we must go beyond leading order in the
Foldy--Wouthuysen reduction, which is not an efficient expansion
at finite density, even though $|{\bf p}|/M$ is quite small.
The problem is that the Foldy--Wouthuysen expansion also involves
$\Phi/M$ and $W/M$, which are less than unity but not negligible.

Reinhard has proposed 
a more useful expansion, which counts powers of a (relatively) small
nucleon velocity $v$ \cite{REINHARD89}.
Specifically,

\begin{eqnarray}
 \epsilon_\alpha/M \equiv (E_\alpha-M)/M
        \qquad   & \longrightarrow & \qquad {\rm order\ } v^2 \ , 
              \nonumber\\[3pt]
  (\Phi-W)/M \qquad & \longrightarrow & \qquad {\rm order\ } v^2 \ ,
                \nonumber\\[3pt]
  \Mstar/M = (M-\Phi)/M \qquad & \longrightarrow & \qquad {\rm order\ } 1 
            \ ,     \nonumber\\[3pt]
 \displaystyle{ {\bbox{\sigma\cdot p}\over M} \,
   {M\over  M+E_\alpha-\Phi-W}\,
      {\bbox{\sigma\cdot p}\over M}} \qquad
        & \longrightarrow &   \qquad {\rm order\ } v^2 \ . 
  \label{eq:orders}
\end{eqnarray}

\noindent
Here $E_\alpha$ is the energy eigenvalue for state $\alpha$, including
the rest mass.
Since nuclei are weakly bound, the kinetic energy proportional to $v^2$
balances the 
net potential energy $\approx \Phi-W$, 
which leads to the first two results.
But while
the difference of the potentials is counted as a small
quantity,  the potentials themselves are not.  
On the other hand, they
are still small enough compared to $M$ to count $\Mstar/M$ as
order 1, which explains the final two counting rules.

Reinhard's procedure constructs order-by-order in $v^2$
a single-particle hamiltonian $\hNR$ that
acts on two-component (normalized) wavefunctions \cite{REINHARD89}.
The latter are related to the upper components of the Dirac four-component
wavefunctions by a normalization operator.
He equates the expectation value  
$\langle \hDirac - E_\alpha \rangle$ calculated with the four-component
wavefunctions
with the corresponding expectation value
$\langle \hNR - \epsilon_\alpha \rangle$ calculated with the two-component
wavefunctions.
This condition specifies $\hNR$ at each order in $v^2$, up to
corrections of the next order.

We have considered two possibilities for 
the spin-orbit part of $\hNR$ to order $v^2$, 
\begin{equation}
   \hLSR = \biggl[ {1\over 2r}{d \over dr}{1\over \Mbar(r)} + O(v^4) \biggr]
       \LdotS
       =  \biggl[ {1\over 4\Mbar^{2}}{1 \over r} 
        \left( {d \Phi \over dr} +  {d W \over dr} \right)
       + O(v^4)\biggr]  \LdotS \ .
     \label{eq:hReinhard}
\end{equation}
where $\Mbar(r)$ is defined as
\begin{equation}
  \Mbar(r) \equiv M - {1\over 2}\Bigl( \Phi(r) + W(r) \Bigr) \ ,
       \label{eq:Mbardef}
\end{equation}
and
\begin{equation}
   \hLSR' = \biggl[ {1\over 2r}{d \over dr}{1\over \Mstar(r)} + O(v^4) \biggr]
       \LdotS
       =  \biggl[ {1\over 2\Mstar{}^{2}}{1 \over r} 
        {d \Phi \over dr}
       + O(v^4)\biggr]  \LdotS \ .
     \label{eq:hpReinhard}
\end{equation}
We have also constructed the $O(v^4)$ corrections to each.
By taking explicit matrix elements of $\hLSR$ and $\hLSR'$ using 
two-component wavefunctions for spin-orbit pairs and comparing to
the energy eigenvalues for the corresponding Dirac wavefunctions, we
can determine the accuracy of these truncations.
We find that $\hLSR$ systematically underpredicts the splittings by
roughly 5\% while $\hLSR'$ overpredicts the splittings by about 15\%.
The predictions using the $O(v^4)$ hamiltonians are quite similar
in the two cases and are within 3\% of the Dirac splittings.
We will use $\hLSR$ to $O(v^2)$ in the sequel. 

We can estimate spin-orbit splittings as follows.
We use the local-density expressions from Eqs.~(\ref{eq:Philda})
and (\ref{eq:Wlda})
in the numerator of Eq.~(\ref{eq:hReinhard}) and replace
$\Mbar$ in the denominator by an average value
\begin{equation}
   \langle \Mbar \rangle = {1\over 2} (M + \langle\Mstar\rangle
         - \langle W \rangle)  \ .  \label{eq:mbarav}
\end{equation}
We use an interpolated value of $\Mstar$:
\begin{equation}
  \langle \Mstar \rangle \equiv 
        \Mstarzero + y (M-\Mstarzero)  \ , \label{eq:interp}
\end{equation}
and Eq.~(\ref{eq:HvH}) to define $\langle W \rangle$ given
$\langle\Mstar\rangle$.
One would expect $y$ to be given roughly by
$y \approx 1 - \langle\rho\rangle/\rhozero$, where $\langle\rho\rangle$
is an average density seen by the nucleons making up the spin-orbit
pair.  
In $^{40}$Ca, this would imply $y \approx 0.2$ for the $p$ states and
$y \approx 0.4$ for the $d$ states.
So the energy splitting takes the form
\begin{equation}
     \epsilon_{{\rm splitting}} =
          C \,{M - \Mbarzero \over \langle \Mbar \rangle^2} \,
            \delta\langle \LdotS \rangle \ ,  \label{eq:split} 
\end{equation} 
where the constant $C$ depends on the average of 
$(1/r)d\rhoB/dr$ over the nucleus (which is essentially the same
for the members of a spin-orbit pair), and
$\delta\langle \LdotS \rangle$ is the difference in
the expectation values of $\LdotS$ for the spin-orbit pair.
The nuclear matter value $\Mbarzero$ follows from $\Mstarzero/M$
using Eqs.~(\ref{eq:Mbardef}) and (\ref{eq:HvH}).

This local-density estimate of the splitting as a function of
nuclear matter $\Mstarzero /M$ is shown in
Figs.~\ref{fig:oxy}--\ref{fig:pb}
as a dotted line.
The curves are normalized to the point with the largest value of 
$\Mstarzero /M$ to absorb the constant $C$ and
any overall normalization errors.
The least-bound states are well fit with $y=0.4$, and the deeply
bound states, with $y=0.2$.
The local-density estimates account quantitatively for the spin-orbit
trend.\footnote{%
Note that fitting to a different charge radius for $^{40}$Ca shifts
the dotted curve up (for a smaller radius) or down (for a larger radius).
That is why the open diamonds tend to be slightly above the
curve, since the $^{40}$Ca
radius used is 3.45\,fm rather than 3.48\,fm as in the other models.}

The shaded regions in Figs.~\ref{fig:oxy}, \ref{fig:cad}, and
\ref{fig:pb} indicate
the empirical spin-orbit splittings. 
Since we do not calculate rearrangement effects,
there are intrinsic uncertainties in
comparing to experimental splittings.%
\footnote{In addition, the Kohn--Sham nature of our calculation
implies that the single-particle energy
eigenvalues do not correspond precisely to observables \cite{SEROT97}. 
Nevertheless, we expect that {\it differences\/}
in eigenvalues are good estimates of the splitting, at least near the
Fermi surface.}
We have therefore been conservative in defining these regions.
However, the intersection of the universal curve with each region
implies that models with effective masses outside the
range $0.58 \leq \Mstarzero/M \leq 0.64$ proposed in Ref.~\cite{FURNSTAHL95}
will not be able to reproduce empirical spin-orbit splittings
(when there is no tensor coupling).
This is true of the QMC \cite{SAITO96} and ZM \cite{BIRO97} models, 
which are indicated by triangles in Figs.~\ref{fig:oxy}--\ref{fig:pb}.

One of the appealing features of relativistic models is that one
can come to a similar conclusion about $\Mstarzero/M$ without considering
spin-orbit splittings.
For example, the study by   
Rufa et al. \cite{RUFA88} of conventional QHD--I models (i.e., the linear
Walecka
model plus cubic and quartic scalar meson self-couplings) revealed that
a good fit to bulk nuclear properties {\it excluding spin-orbit splittings\/}
still required $\Mstarzero/M$ to be close to $0.6$ at equilibrium.

This result can be understood in terms of 
a key feature of relativistic models: the additional
saturation mechanism that arises from a Lorentz scalar interaction.
The velocity dependence in the interaction due to the
scalar potential appears in the energy of nuclear matter 
(to leading order in $k^2$) as a modification of the kinetic energy
per particle: 
\begin{equation}
  \left\langle {k^2\over 2 M}\right\rangle  \longrightarrow
  \left\langle {k^2\over 2 \Mstar}\right\rangle
  = {3 \over 5}{\kfermi^2\over 2M} \left({M\over \Mstar(\rhoB)}\right) 
     \ .  \label{eq:nm}
\end{equation}
The extra density-dependent repulsion from the $M/\Mstar$ factor
leads to saturation \cite{SEROT97}; 
in successful mean-field models 
(which include QHD--I models with $0.58 \leq \Mstarzero/M \leq 0.64$)
it is the dominant saturation mechanism.
In these cases, a fit to
the nuclear matter equilibrium density and binding energy alone constrains
$\Mstarzero/M$ and the implied velocity dependence; the four-component
structure of the Dirac wave functions then {\em automatically\/} generates
an appropriate spin-orbit force in finite nuclei.
In contrast, in QHD--I models with larger values of $\Mstarzero/M$, the
empirical equilibrium point is obtained from repulsive three- and
four-body contributions that become more and more important as
$\Mstarzero/M$ increases.

In summary, models without isoscalar tensor couplings exhibit a strong
correlation between the effective nucleon mass at nuclear matter
equilibrium density and the spin-orbit splittings in nuclei.
Comparison to experiment reinforces the constraint 
$0.58 \leq \Mstarzero/M \leq 0.64$ for successful models of this type.
Next we consider the generalization that includes tensor couplings.

\section{Tensor coupling and spin-orbit splitting}
\label{sect:three}

We generalize $\hDirac$ from Eq.~(\ref{eq:hdirac}) by adding
a tensor potential $T(r)$:
\begin{equation}
   \hTDirac = -i\bbox{\nabla\cdot\alpha} + 
        \beta \Bigl(M-\Phi(r)\Bigr)
      +  W(r) - i\beta\bbox{\alpha\cdot\widehat r}\,
       T(r) \ . \label{eq:hdiracp}
\end{equation}
A tensor potential of this form arises in a meson model from the tensor
coupling of an isoscalar vector meson ($\omega$) to the nucleon.
Using the conventions of Ref.~\cite{FURNSTAHL97},
\begin{equation}
   {\cal L}_{{\rm tensor}} =
      - {\fv\gv\over 4M}\overline N\sigma^{\mu\nu}N \,
      (\partial_\mu V_\nu - \partial_\nu V_\mu)
                       \ , \label{eq:Lvmdt}
\end{equation}
which implies
\begin{equation}
   T(r) = {\fv\over 2M} {dW\over dr}  \ .  \label{eq:Tvmd}
\end{equation}
(Note that $\fv$ as defined here corresponds to the combination
$\fv/\gv$ typically appearing in one-boson-exchange models
\cite{MACHLEIDT}.)
The analog of Eq.~(\ref{eq:Lvmdt}) in the point-coupling model
of Ref.~\cite{RUSNAK97} is
\begin{equation}
   {\cal L}_{{\rm tensor}} =
      - {\fvt\over \fpi^2\Lambda}      
      \overline N\sigma^{\mu\nu} N \,
      \partial_\mu (\overline N \gamma_\nu N)
                       \ , \label{eq:Lpct}
\end{equation}
which implies
\begin{equation}
   T(r) = {\fvt\over \fpi^2\Lambda} {d\rhoB\over dr}  \ .  \label{eq:Tpc}
\end{equation}
Thus we can make the rough correspondence $\fv \leftrightarrow 2\fvt$.
[Note that a spin-2 meson in the (1,1) representation is described by a
{\em symmetric\/} tensor $f^{\mu\nu}$.
Thus it does not couple to $\overline N\sigma^{\mu\nu} N$ and will not
produce a tensor potential at the mean-field level.]

We repeat the procedure that led to
Eq.~(\ref{eq:hReinhard}) to find
\begin{equation}
   \hLST =   \left[ {1\over 4\Mbar^{2}} 
        {1 \over r}\left({d \Phi \over dr} + {d W \over dr} \right)
        + {\fv \over 2 M \Mbar} {1 \over r}{d W \over dr}
         + O(v^4)   \right]  \LdotS \ ,
     \label{eq:hLStensor}
\end{equation}
and the result for point-coupling models is found from the substitution
\begin{equation}
  {\fv \over M}{d W \over dr} \longrightarrow {2\fvt \over \Lambda \fpi^2}
         {d \rhoB \over dr}  \ .  \label{eq:hLStensort}
\end{equation}
Now suppose that we want to know how  $\Mstar$ in a model with
$\fv \neq 0$ compares to the value $\Mstar_{\fv=0}$ in
an ``equivalent'' model with $\fv = 0$.  If we define ``equivalent''
to mean that they have the same spin-orbit strength, we can use
Eq.~(\ref{eq:hLStensor}) to make a rough comparison.  Let us assume
the local-density relations from Eqs.~(\ref{eq:Philda}) and (\ref{eq:Wlda}),
so that if we equate $\hLST$ in the two models, the $r$ dependence
cancels (in an averaged sense), and we have the condition
\begin{equation}
  \left.{M - \langle \Mbar\rangle 
          \over \langle\Mbar\rangle^2}\right|_{\fv=0} =
     {M - \langle\Mbar\rangle \over \langle\Mbar\rangle^2}
     + {\fv\langle W\rangle \over M\langle\Mbar\rangle} \ . 
     \label{eq:equiv}
\end{equation}
The analogous condition for the point-coupling models follows after
applying Eq.~(\ref{eq:hLStensort}). 

We can use Eqs.~(\ref{eq:equiv}) and (\ref{eq:HvH}) to transform the region
$0.58 \alt (\Mstarzero/M)_{\fv=0} \alt 0.64$, which applies to $\fv=0$,
to the ``equivalent'' region for $\fv >0$ or $\fvt > 0$.
Values of $\langle\Mbar\rangle$ and $\langle\Mbar\rangle_{\fv=0}$ 
are defined by Eq.~(\ref{eq:mbarav}) in terms of $\Mstarzero$ and
$(\Mstarzero)_{\fv=0}$ using Eq.~(\ref{eq:interp})
with $y=0.2$,
and then $\langle W \rangle$ and $\langle W \rangle_{\fv=0}$ 
follow from Eq.~(\ref{eq:HvH}). 
One then inverts the resulting relation for $\Mstarzero$ given an input
$(\Mstarzero)_{\fv=0}$.

\begin{figure}[p]
 \setlength{\epsfxsize}{3.8in}
  \centerline{\epsffile{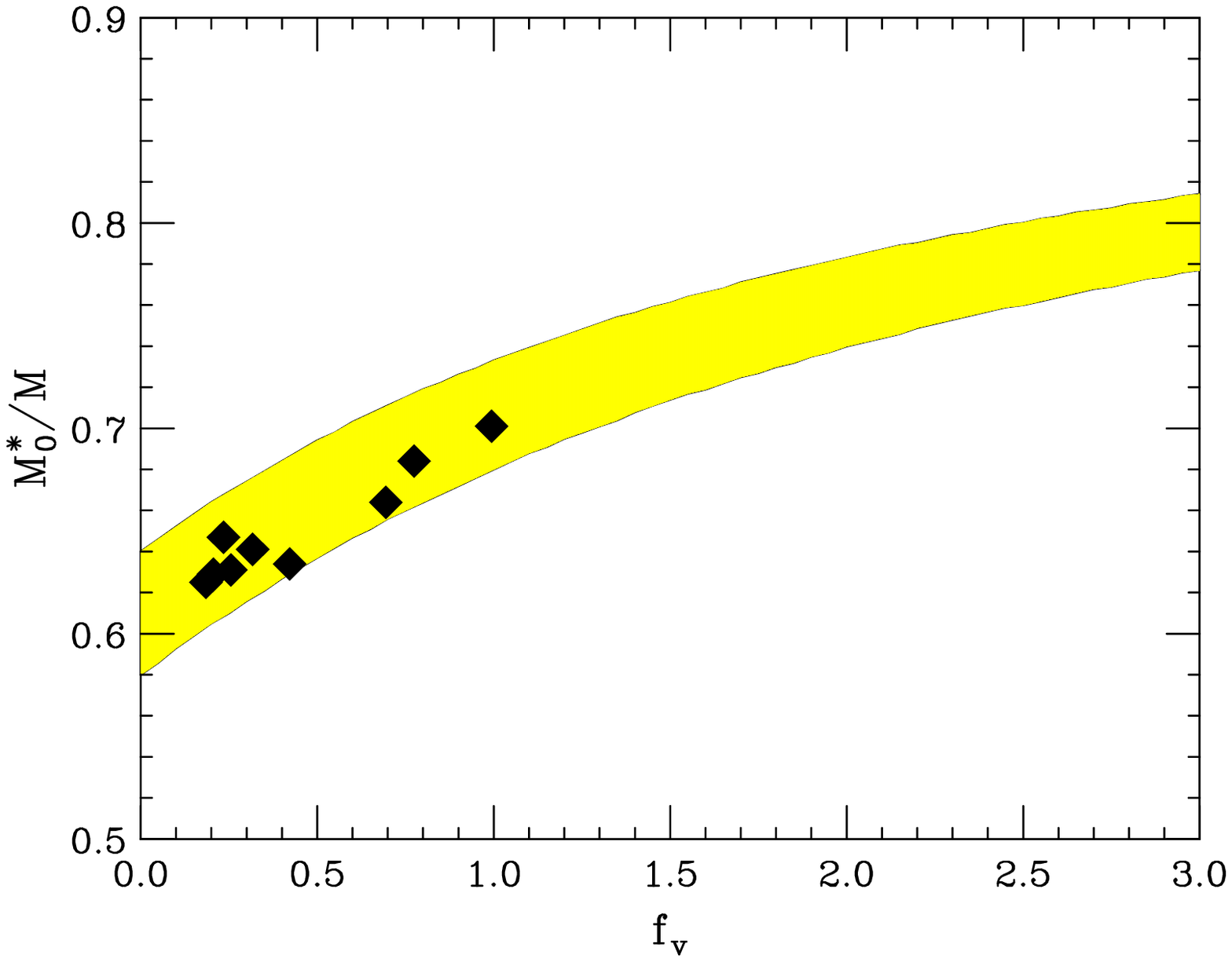}}
\vspace{.1in} \caption{\small Correlation between the isoscalar
tensor coupling $\fv$ and the effective mass in successful 
mean-field models with mesonic degrees of freedom.
The shaded region shows the range $\Mstarzero/M$ for a given
value of $\fv$ that corresponds to
the range $0.58 \leq \Mstarzero/M \leq 0.64$ with $\fv=0$, 
found using Eq.~(\protect\ref{eq:equiv}).
The QHD models, shown as filled diamonds, 
are direct fits to nuclear observables
from Refs.~\protect\cite{FURNSTAHL97} and
\protect\cite{RUSNAK97b}.}
 \label{fig:one}
\end{figure}

\begin{figure}[p]
 \setlength{\epsfxsize}{3.8in}
  \centerline{\epsffile{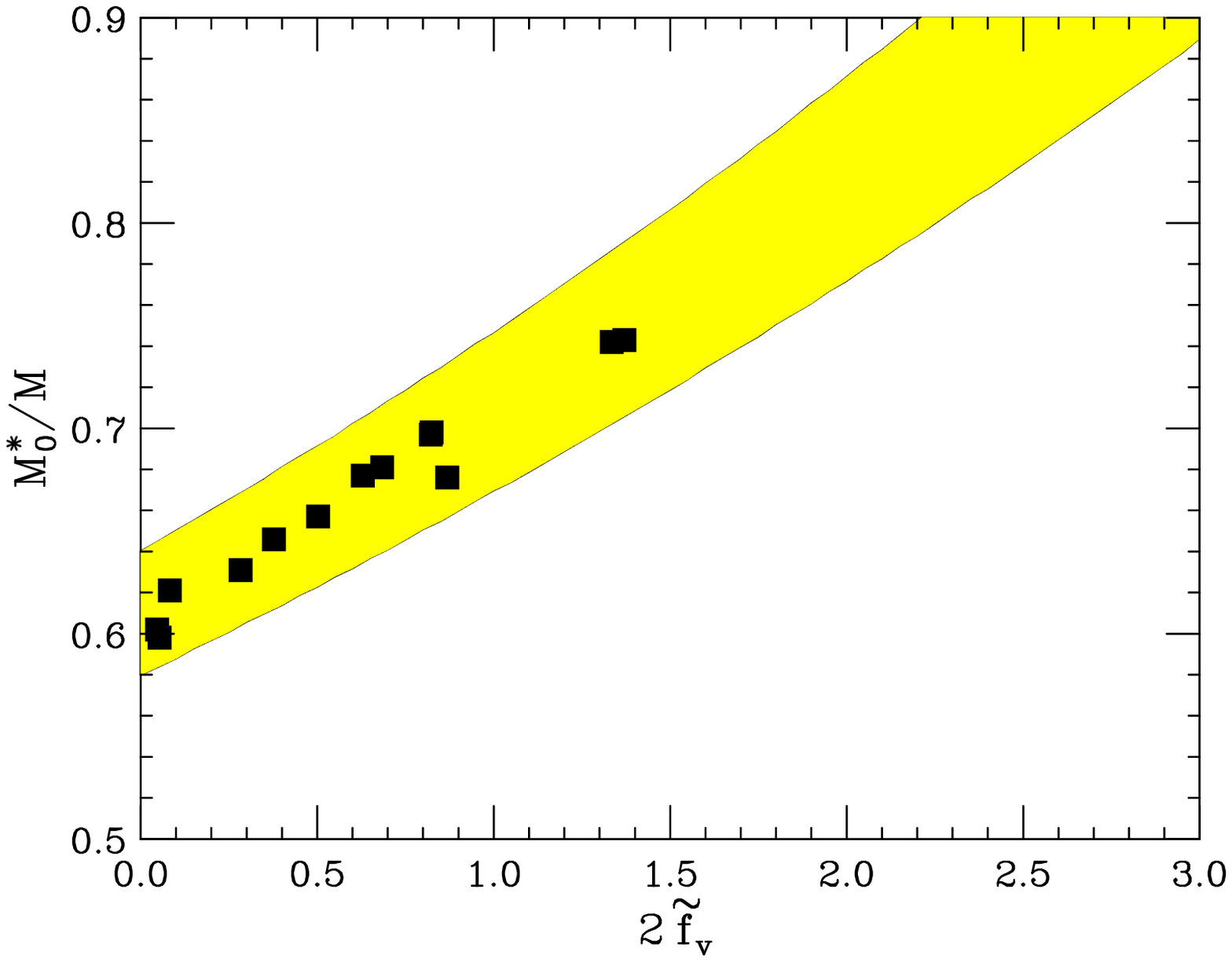}}
\vspace{.1in} 
\caption{\small Correlation between the isoscalar
tensor coupling $\fvt$ and the effective mass in successful 
mean-field point-coupling models.
The shaded region shows the range $\Mstarzero/M$ for a given
value of $\fvt$ that corresponds to
the range $0.58 \leq \Mstarzero/M \leq 0.64$ with $\fvt=0$, 
found using Eq.~(\protect\ref{eq:equiv}).
The point-coupling models, shown as filled squares, 
are direct fits to nuclear observables
from Ref.~\protect\cite{RUSNAK97}.}
 \label{fig:two}
\end{figure}

The transformed regions are shown in Figs.~\ref{fig:one} and \ref{fig:two},
along with diamonds and squares marking the generalized ($\fv \not= 0$)
models from Refs.~\cite{FURNSTAHL97}, \cite{RUSNAK97} and \cite{RUSNAK97b}.
While the shaded regions are not definitive indicators of
the relation between the tensor coupling and $\Mstar$ in successful models,
they should provide a reliable estimate of the basic trend.
Evidently, the bands are consistent with the models with nonzero
tensor coupling that were fit directly to nuclei.
These results are also consistent with new fits in the point-coupling
models from Ref.~\cite{RUSNAK97}, in which $\fvt$ is
held fixed at zero.  The new sets have
$(\Mstarzero/M)_{\fv=0}$ between 0.60 and 0.64.

As part of an extensive optimization analysis of conventional QHD
models, Rufa et al.\ considered including an isoscalar tensor
coupling \cite{RUFA88}.
They noted the trade-off between the magnitude of the spin-orbit
splitting and the size of the scalar field.
However, their chi-squared optimization yielded only
a small improvement from the new term,%
\footnote{The larger values of $\fv$ and $\fvt$ from optimized fits in 
Refs.~\cite{FURNSTAHL97,RUSNAK97b,RUSNAK97} also yielded only a small
improvement in $\chi^2$ (a reduction of about one out of forty)
compared to similar models with these parameters
fixed at zero.} 
and the best value had a small value of $\fv \approx 0.3$.
(Note that their conventions differ from ours by a minus sign.)
Reinhard concludes that including the tensor
would make a difference only in the details, but not in the
qualitative picture \cite{REINHARD89}.

In Ref.~\cite{BIRO97}, Bir\'o and Zim\'anyi propose adding a tensor coupling
of the $\omega$ meson to the nucleon to the lagrangian proposed by
Zim\'anyi and Moszkowski \cite{ZIMANYI90}, 
in order to bring the spin-orbit force up
to the strength implied by nuclear spin-orbit splittings.
They perform a similar expansion to the one above.
However, the value of $\Mstarzero/M$ in the ZM model is roughly 0.85.
Figure~\ref{fig:one} shows that this would require a very large value
of $\fv$, which does not appear to be compatible with good fits to nuclear
observables.

What constraints exist on the value of $\fv$?
In free space there are constraints from nucleon 
electromagnetic form factors under the assumption of vector dominance.
Depending on the form of vector dominance (i.e., tensor or vector
representation \cite{BORASOY96}), we obtain a limit on $\fv$
from the isoscalar anomalous moment of the nucleon
or from the isoscalar magnetic radius of the nucleon.
In either case, one finds the small value 
$| \fv | \alt 0.2$ \cite{HOHLER76,BROWN86,MEISSNER87,MEISSNER}.
From Fig.~\ref{fig:one}, we see that such a limit implies an unimportant
change in $\Mstar$ from the results with $\fv = 0$.
In one-boson-exchange potentials, the {\em isovector\/} tensor coupling is
important, but the isoscalar tensor coupling is usually taken
to be zero \cite{MACHLEIDT}.
On the other hand, without further assumptions, the value of $\fvt$ in
the point-coupling models is not related to the electromagnetic structure
of the nucleon and is therefore unconstrained.

It is important to remember, however, that
the mean-field parameters obtained from fitting to finite
nuclei can be very different from the
``free-space'' tree-level values.
These mean-field
parameters must incorporate the effects of many-body physics beyond
the Hartree level, which could induce strong renormalizations.
We anticipate that isovector couplings will be less affected (since the
free-space values are already of natural size and nuclei
have roughly equal numbers of neutrons and protons), and therefore
argue that fitting isovector parameters using free-space electromagnetic 
form factors is a good approximation.  
However, the situation with isoscalar parameters is unclear.
We might even expect that in the absence of symmetry constraints in
the medium, they take natural values.  This would mean that any
values shown in Figs.~\ref{fig:one} or \ref{fig:two} are accessible 
(here, as in Ref.~\cite{FURNSTAHL97}, $\fv/4$ is the natural combination).

One might hope to constrain the value of $\fv$ from the energy dependence
of the optical potential.
A single-particle hamiltonian of the form $\hDirac$ in Eq.~(\ref{eq:hdirac})
implies an energy-dependent optical potential for nucleon-nucleus 
scattering of the form \cite{SEROT86}
\begin{equation}
  \Vopt(r;\Phi,W) = 2 E W(r) - W(r)^2 + \Mstar(r)^2 - M^2  \ .
    \label{eq:Vopt}
\end{equation}
If the potential $\Vopt(r)$ follows the nuclear density, the slope
in energy is given by $2\Wzero$ [see Eq.~(\ref{eq:Wlda})],
and is constrained by low-energy
neutron-nucleus scattering to be about $0.6\,M$\cite{SIEMENS87,SEROT86}.
To be consistent with experiment, $\Wzero$ must be as large as the
values implied by $0.58 \leq \Mstarzero/M \leq 0.64$
through Eq.~(\ref{eq:HvH}).
The larger values of $\Mstarzero$, and consequently smaller
values of $\Wzero$, when
$\fv >0$ (or $\fvt > 0$) could set a limit on the magnitude
of the tensor coupling.

We cannot identify $\Vopt$ directly from $\hTDirac$ because of the
tensor potential.
However, we can apply field redefinitions to the
nucleon wavefunctions, 
so that the transformed version of Eq.~(\ref{eq:hdiracp})
does not contain a tensor potential.
Such a transformation was applied to Dirac wavefunctions in
Ref.~\cite{CLARK85}.
Given a solution $\psi({\bf r})$ that satisfies
$\hTDirac\psi({\bf r}) = E \psi({\bf r})$
for scattering states ($E>M$), 
we define
a transformed wavefunction $\phi({\bf r})$ by
\begin{equation}
  \phi({\bf r}) = e^{-U(r) \gammazero} \psi({\bf r}) \ ,
    \label{eq:transphi}
\end{equation} 
where $U(r) = \fv W(r)/2M$ in the meson model 
and $U(r) = \fvt \rhoB(r)/\fpi^2\Lambda$
in the point-coupling model.
The function $U(r)$ goes to zero rapidly outside the nucleus,
so scattering observables are unchanged.

The new wavefunction $\phi$ satisfies
\begin{equation}
  \Bigl[ -i \bbox{\alpha\cdot\nabla} + W e^{2U\gammazero}
    + \beta (M-\Phi) e^{2 U\gammazero} \Bigr]
   \phi_\alpha({\bf r}) = E e^{2U\gammazero} \phi({\bf r}) \ .
        \label{eq:newphi}
\end{equation}
By rewriting the exponents we can make the identifications:
\begin{eqnarray}
  \Mstar_{\fv=0} &=& \Mstar \cosh 2U
           + ( W - E) \sinh 2U
           \ , \label{eq:mstara} \\
   W_{\fv=0}  &=&
     E\ + ( W - E)
       \cosh 2U + \Mstar \sinh 2U
          \ .  \label{eq:Wa}
\end{eqnarray}
Therefore the transformation introduces energy dependence into
the effective scalar and vector potentials.
If evaluated in nuclear matter for $E \approx \mu$, the
{\em predicted\/} values of 
$(\Mstarzero)_{\fvt=0}$ and $(\Wzero)_{\fvt=0}$ for the
point-coupling models with $\fvt \neq 0$ are
consistent with Fig.~\ref{fig:two};
in particular, one obtains 
$0.58 \leq (\Mstarzero/M)_{\fvt=0} \leq 0.61$ for each model shown.
The predicted values for the QHD models in Fig.~\ref{fig:one},
however, are slightly smaller than
expected [$(\Mstarzero/M)_{\fv=0}$ is roughly 0.55].

If we use $\Mstar_{\fv=0}$ and $W_{\fv=0}$ from Eqs.~(\ref{eq:mstara}) and
(\ref{eq:Wa}) to evaluate the optical
potential in Eq.~(\ref{eq:Vopt}), we find
\begin{equation}
  \Vopt(r;\Phi_{\fv=0},W_{\fv=0}) = \Vopt(r;\Phi,W) \ .
    \label{eq:Voptb}
\end{equation}
That is, the additional energy dependence in Eqs.~(\ref{eq:mstara}) and
(\ref{eq:Wa}) cancels out when evaluating $\Vopt$.
Thus the energy dependence of the optical potential 
is simply proportional to 
the {\it untransformed\/} vector potential $W$ and is 
{\em independent\/} of $\fv$ (or $\fvt$).
If the tensor coupling is too large,
Figures~\ref{fig:one} and \ref{fig:two} 
imply that $\Wzero$ will be too small to account
for the experimental energy dependence.
Therefore, smaller values of $\fv$ (or $\fvt$) are favored, with an
upper limit around unity.
This analysis is not conclusive, however, since the radial dependence
of the potentials may invalidate the simple argument leading to
the comparison with experiment \cite{SEROT86}.
This issue will be considered in future work.

\section{Summary}

We examined the role of an isoscalar tensor coupling to the
nucleon in chiral effective lagrangians for nuclei.
Effective lagrangians must contain all nonredundant terms consistent with
the underlying symmetries.
Since tensor couplings of the vector mesons
to the nucleon (or their analogs in point-coupling models) will
appear at low order in an expansion in powers of fields (or densities) and
derivatives, they should be included.

In models that ignore such couplings, there is a strong correlation between
the value of the nucleon effective mass $\Mstarzero$ at equilibrium nuclear
matter density and spin-orbit splittings in nuclei.
In fact, given a value of $\Mstarzero$, one can predict the {\it calculated\/}
splittings of weakly bound levels to an accuracy of roughly 0.3\,MeV.
Moreover, to accurately reproduce the {\it empirical\/} splittings, 
one requires an equilibrium
$\Mstarzero / M$ between 0.58 and 0.64.
Interestingly, similar values of $\Mstarzero$ are obtained when fits are
performed without including information from these splittings.

When an isoscalar tensor coupling is included and allowed to vary freely,
fits to nuclear properties slightly
favor natural-sized values of the coupling,
which are significantly larger than values deduced from free-space
electromagnetic observables.
This is particularly true in point-coupling models.
These larger values of the coupling modify the simple relationship between
$\Mstarzero$ and the spin-orbit splittings noted above; however,
it is possible to estimate the contribution of the tensor interaction to
the spin-orbit force and to show that the larger values of $\Mstarzero$
obtained in nuclear matter are consistent with the observed splittings.

We emphasize that the small size of the isoscalar tensor coupling deduced
from free-space electromagnetic data is not incompatible with the larger
values obtained in fits to nuclei, since the in-medium parameters must
contain many-body effects.
In all of the models discussed here, the resulting tensor coupling is
still natural.
It remains to be seen whether explicit calculation of the many-body
contributions can account for the larger values favored by nuclei.
Moreover, it may be possible to constrain the tensor coupling by including
additional observables in the fits; the energy dependence of the
optical potential for nucleon-nucleus scattering is a promising candidate.

\acknowledgments

We thank B.~C.~Clark for useful discussions.
This work was supported in part by the Department of Energy
under Contracts No.\ DE--FG02--87ER40365 and DE--FG02--87ER40328, 
and the 
National Science Foundation
under Grants No.\ PHY--9511923 and PHY--9258270.
We thank the Department of Energy's Institute for Nuclear Theory
at the University of Washington for its hospitality and the
Department of Energy for partial support during the completion
of this work.


\end{document}